\documentclass[a4paper,11pt]{article}
\usepackage{pos}

\usepackage{slashed}
\usepackage{multirow}
\usepackage{dcolumn}
\newcommand\Tstrut{\rule{0pt}{2.6ex}}         

\def\alambda{A_\lambda}
\def\akappa{A_\kappa}
\def\mueff{\mu_\mathrm{eff}}

\def\tanb{\tan\beta}

\def\sQ3{\widetilde{Q}_3}
\def\sU3{\widetilde{U}_3}
\def\sD3{\widetilde{D}_3}
\def\bino{\widetilde{B}}
\def\wino{\widetilde{W}}

\def\higgsinod{\widetilde{H}^0_d}
\def\higgsinou{\widetilde{H}^0_u}
\def\singlino{\widetilde{S}}
\def\mhsm{m_{h_{\mathrm{SM}}}}

\newcommand{\vu}{v_u}
\newcommand{\vd}{v_d}
\newcommand{\vs}{v_s}
\def\vevs{{\it vevs}}
\def\vu{v_u}
\def\vd{v_d}

\def\vs{v_{\!_S}}

\def\beq{\begin{equation}}
\def\eeq{\end{equation}}
\def\beqa{\begin{eqnarray}}
\def\eeqa{\end{eqnarray}}

\allowdisplaybreaks
\newcommand{\fbinv}{\text{fb}$^{-1}$}

\def\nmssmtools{{\tt NMSSMTools}}
\def\checkmate{{\tt CheckMATE}}
\def\smodels{{\tt SModelS}}

\def\higgsbounds{{\tt HiggsBounds}}
\def\higgssignals{{\tt HiggsSignals}}

\def\pythia8{{\tt PYTHIA8}}

\def\cosmotransitions{{\tt CosmoTransitions}}
\def\z3nmssm{$Z_3$-NMSSM}
\newcommand{\bea}{\begin{eqnarray}}
\newcommand{\eea}{\end{eqnarray}}

\title{Prospects of a Strong first-order Electroweak Phase Transition in the $Z_3$-NMSSM}

\author[a]{Arindam Chatterjee}
\author[b]{AseshK Datta}
\author*[b,c]{Subhojit Roy}

\affiliation[a]{Department of Physics, School of Natural Sciences, Shiv Nadar University, Gautam Budhha Nagar, Uttar Pradesh 201314, India}
\affiliation[b]{Harish-Chandra Research Institute, A CI of Homi Bhabha National
Institute, Chhatnag Road, Jhunsi, Prayagraj (Allahabad) 211019, India}
\affiliation[c]{Regional Centre for Accelerator-based Particle Physics, Harish-Chandra Research Institute, \\ Prayagraj (Allahabad) 211019, India}

\emailAdd{arindam.chatterjee@snu.edu.in}
\emailAdd{asesh@hri.res.in}
\emailAdd{subhojitroy@hri.res.in}

\abstract{We have studied the viability and possible patterns of a strong
first-order electroweak phase transition (SFOEWPT) within the $Z_3$-symmetric Next-to-Minimal Supersymmetric 
Standard Model (NMSSM), in view of the latest experimental results from the dark matter (DM) sector, Higgs sector and the searches of the lighter chargino and neutralinos at the Large Hadron Collider (LHC). We show that the region of parameter space with relatively small $\mu_\mathrm{eff}$ that favors an SFOEWPT has started to get excluded from the searches at the LHC and various DM experiments. However, there still remain phenomenologically much involved and compatible regions for an SFOEWPT that are yet not 
sensitive to the latest LHC and DM searches. We further estimate the production of stochastic gravitational waves (GW) from an SFOEWPT within and without the bag model and the prospects of detecting such signals at various future/proposed GW experiments.}

\FullConference{%
  41st International Conference on High Energy physics - ICHEP2022\\
  6-13 July, 2022\\
  Bologna, Italy
}

\begin{document}
\maketitle
\section{Introduction}
\vspace{-2mm}
Electroweak Baryogenesis (EWBG) is the process giving rise to the observed baryon asymmetry of the Universe (BAU) at the
 electroweak scale.
This process requires  a strong first-order phase transition (FOPT) to occur during the electroweak phase transition (EWPT) to met one of the well-known Sakharov's conditions, that is, the system has to go out of equilibrium.
However, such an FOPT (and hence EWBG) is not possible in the Standard Model (SM) which can be traced to the fact that the experimentally observed SM-like Higgs boson has been found to be too heavy ($\mhsm \approx 125$ GeV) for the purpose.
It is well known that popular supersymmetry (SUSY) extensions of the SM, viz., the 
Minimal SUSY SM (MSSM), its next-to-minimal 
version (NMSSM), etc., could provide a favorable setup for EWBG to take place.
An effective Higgs potential that favours an FOPT in the early Universe can be found in the  presence of an extended Higgs sector and other scalar degrees of freedom (in particular, the top squarks) of these scenarios.
However, null results from the searches for suitable light top squarks at the LHC have made the situation extremely tight for an strong first-order electroweak phase transition (SFOEWPT) to take place in the MSSM.
 On the other hand, the presence of a gauge singlet scalar field makes the NMSSM more favorable for SFOEWPT.

In this paper, we examine the possibilities of an SFOEWPT in the $Z_3$-symmetric NMSSM ($Z_3$-NMSSM) after meeting the latest constraints from the LHC and the DM sector.
The effective higgsino mass parameter ($\mueff$) is the most important parameter of the present model that connects the phenomenology of the DM, the LHC and the EWPT sectors. This is since apart from $\mueff$ there are a few other model parameters that appear both in the Higgs and the  electroweakino sectors.
We will find that such a connection raises the tantalising possibility that relatively light higgsinos with masses below a few hundred GeV, in the presence of light singlet-like scalars and a lighter singlino and/or a bino, can somewhat comfortably escape the latest LHC constraints from their searches~\cite{Abdallah:2020yag, Datta:2022bvg}.
In particular, we would like to explore how small a $\mueff$ might still be feasible in the light of the latest experimental constraints given that an SFOEWPT prefers the same and which is also motivated by `naturalness'.
In addition to this, stochastic gravitational waves (GW) can be generated from the dynamics of the nucleated bubbles during an SFOEWPT.
 In this work, we estimate the GW spectra for a chosen set of benchmark scenarios and study the prospects of detecting GWs in future proposed experiments. 

The present work is organized as follows.
We provide a short description of the $Z_3$-NMSSM scenario in section~\ref{sec:model}.
We discuss our results in section~\ref{results} where we present a few benchmark scenarios that exhibit SFOEWPT but are excluded from the LHC searches a for the electroweakinos. Later on we demonstrate few scenarios that pass latest constraints from various experiments. We discuss how SFOEWPT is 
realized even for a scenarios with a relatively small $\mueff$.
Finally, we present the  prospects of detecting the GW in future proposed GW 
experiments in these viable benchmark scenarios. 
In section~\ref{sec:summary} we provide an overview of our present work and a forecast for the future.  The details of this work can be found in reference~\cite{Chatterjee:2022pxf}.
\section{The theoretical framework of the $Z_3$-NMSSM} 
\label{sec:model}
The superpotential of the $R$-parity conserving $Z_3$-NMSSM is given by
\beq
{\cal W}= {\cal W}_\mathrm{MSSM}|_{\mu=0} + \lambda \widehat{S}
\widehat{H}_u \cdot \widehat{H}_d
        + {\kappa \over 3} \widehat{S}^3 \, ,
\label{eqn:superpot}
\eeq
where ${\cal W}_\mathrm{MSSM}|_{\mu=0}$ is the MSSM superpotential without the
higgsino mass term (the $\mu$-term). $\widehat{H}_u, \widehat{H}_d$ and 
$\widehat{S}$ denote the $SU(2)$ Higgs doublet superfields and the gauge singlet 
superfield, respectively. The (real) scalar field of the singlet superfield $\widehat{S}$ 
develops a non-zero vacuum expectation value ({\it vev}), $\vs$, during EWPT which 
generates an effective $\mu$-term as $\mu_{\rm eff}=\lambda \vs/\sqrt{2}$.
The soft SUSY-breaking Lagrangian is given by
\beq
-\mathcal{L}^{\rm soft}= -\mathcal{L_{\rm MSSM}^{\rm soft}}|_{B\mu=0}+ m_{S}^2
|S|^2 + (
\lambda A_{\lambda} S H_u\cdot H_d
+ \frac{\kappa}{3}  A_{\kappa} S^3 + {\rm h.c.}) \,,
\label{eqn:lagrangian}
\eeq
where $m_S$ denotes the soft SUSY-breaking mass of the singlet scalar field, `$S$', 
$\alambda$ and
$\akappa$ are trilinear soft parameters with mass dimensions one. 
These complex scalar fields can be expressed as 
\begin{align}
H_u= \begin{pmatrix} H_u^+\\ \tfrac{1}{\sqrt{2}} \left(h_u + i a_u\right) 
\end{pmatrix}, \quad \quad
H_d= \begin{pmatrix} \tfrac{1}{\sqrt{2}}  \left(h_d + i a_d\right) \\ H_d^- 
\end{pmatrix}, \quad \quad
S= \frac{1}{\sqrt{2}} \left(s + i \sigma\right),
\label{complexfields}
\end{align}
where $\langle h_u\rangle= v_u$, $\langle h_d\rangle= v_d$ and
$\langle s\rangle= v_s$ are the \vevs~ of the real components
($CP$-even) of the neutral scalar fields at zero temperature. Note that $\sqrt{\vu^2 + \vd^2}= v \simeq 246$ 
GeV with $\tanb= \vu/\vd$.
The present model consists of three $CP$-even, two 
$CP$-odd and two charged physical Higgs states. 
The neutralino sector of the $Z_3$-symmetric NMSSM consists of five neutralinos
which are mixtures of bino ($\bino$), wino ($\wino^0_3$), two higgsinos 
($\higgsinod$, $\higgsinou$) and a singlino ($\singlino$), the last one being the 
fermionic component of the singlet superfield $\widehat{S}$ appearing in the 
superpotential of equation~\ref{eqn:superpot}. The chargino sector of the $Z_3$-symmetric NMSSM and MSSM is exactly similar.

\section{Results}
\label{results}
The latest theoretical and experimental constraints eliminate a certain region of parameter space that exhibit SFOEWPT. 
The theoretical constraints ensure the allowed region of parameter space free from the 
unphysical global minimum of the scalar potential, tachyonic states and the evolutions of various model couplings  with energy do not encounter Landau poles, etc.
The experimental constraints include those coming from the hunt for new physics at the colliders, as well as from the Higgs, DM and flavour sectors.
Further, we ensure that after the SFOEWPT, which facilitates EWBG, the system finally ends up in the physical electroweak minimum at zero temperature. In this work, we use publicly available packages like 
\nmssmtools~{\tt (v5.5.3)}, \checkmate~{\tt (v2.0.34)}, \smodels~{\tt (v2.1.1)}, \higgsbounds~{\tt (v5.8.0)}, \higgssignals~{\tt (v2.5.0)}, and \cosmotransitions~{\tt (v2.0.6)} for our numerical analysis and take into account all the above mentioned constraints. 

We present a few benchmark scenarios in which a SFOEWPT is realized and which satisfy all other theoretical and experimental
constraints but are excluded from the searches of the lighter electroweakinos at 
the LHC.
A small value of $\mueff$ that prefers SFOEWPT draws a significant constraints from the latest DMDD and LHC searches~\cite{Abdallah:2020yag}.
Our goal has been to identify the smallest possible values of $\mueff$ (except for the scenarios when the higgsino-like neutralino/chargino is the lightest of all the electroweakinos) that are  allowed under different 
circumstances. We also present a few allowed scenarios with optimally light higgsinos that satisfies all the constraints and favor an SFOEWPT.
%
\begin{table}[tp]
\renewcommand{\arraystretch}{1.11}
\begin{center}
{\tiny\fontsize{6.3}{6.4}\selectfont{
\begin{tabular}{|c|@{\hspace{0.08cm}}c@{\hspace{0.08cm}}|@{\hspace{0.08cm}}c@{\hspace{0.08cm}}|c|@{\hspace{0.08cm}}c@{\hspace{0.08cm}}|c|}
\hline
\Tstrut
Inputs/Observables & {BP-D1} & {BP-D2} & {BP-D3}  \\
\hline
\Tstrut
$\lambda, \, \kappa, \, \tan\beta$   &  $0.683, \, 0.060, \, 4.77 $ 
                      &  $0.547, \, 0.044, \, 2.87$
                      &  $0.565, \, 0.071, \, 2.87$ \\
$A_\lambda, \, \akappa, \, \mueff, \, M_1$~(GeV)  &  $-1352.3, \, 134.5, \, -274.4, \, 478.8$ 
                               &  $978.4,   \, -110.0, \, 308.0, \, 460.3$
                               &  $963.5, \, -112.5, \, 308.0, \, -57.2$ \\
$m_{\widetilde{Q}_3}, \, m_{\widetilde{U}_3, A_{t}}$~(GeV)  
   &  2956.7, \, 3378.3, \, $-1019.7$
   &  3710.8, \, 3562.8, \, 2204.0
   &  3710.8, 3562.8, \, 2204.0 \\
\hline
\Tstrut
$m_{\chi_{1,2,3,4}^0, \chi_1^\pm}$~(GeV)  &  60.9, -304.3, 307.9, 479.4, -284.1 
                              &  60.6, 312.7, $-338.3$, 468.1, 316.3 
                              &  $-59.6$, 91.1, 327.2, $-338.4$, 316.3 \\
$m_{h_1}, \, m_{h_2}, \, m_{a_1}, \, m_{H^{\pm}}$~(GeV)  &  79.2, 124.4, 126.6, 1359.0 
                                         &  78.1, 122.2, 109.5, 963.8
                                         &  86.9, 123.0, 142.6, 963.6  \\
\Tstrut
$\Omega h^2$  &  $4.9\times10^{-4}$  &  $4.4\times10^{-4}$ & $4.8 \times10^{-3}$ \\[0.10cm]
$\sigma^{\rm SI}_{\chi^0_1-p(n)}\times \xi$~(cm$^2$)  &  $4.5(4.6)\times 10^{-47}$  &  $2.4(2.5)\times 10^{-47}$ &  $2.5(2.6)\times 10^{-47}$\\[0.30cm]
$\sigma^{\rm SD}_{\chi^0_1-p(n)}\times \xi$~(cm$^2$)  &  $3.5(3.2)\times 10^{-42}$   &  $7.6(5.8)\times 10^{-43}$ &  $1.9(1.5)\times 10^{-43}$\\[0.15cm]
\hline
\Tstrut
\texttt{First $T_c$ {\tt(GeV)} / Transition type}  &  129.4 / 1st-order    & 151.5 / 1st-order & 165.7 / 1st-order \\
\texttt{$\{h_d,h_u,s\}_{_\text{False\_vac.}}$} {\tt (GeV)} & $\{0,~0,~0\}$  &  $\{0,~0,~0\}$ &  $\{0,~0,~0\}$  \\
\texttt{$\{h_d,h_u,s\}_{_\text{True\_vac.}}$} {\tt (GeV)}  & $\{25.5,145.6,-474.4\}$   & $\{0,0,539.9\}$ & $\{0,0,557.5.9\}$ \\ \hline
%
%
\texttt{Second $T_c$ {\tt (GeV)} / Transition type}  &  $-$  & 112.7 / 2nd-order & 105.6 / 1st-order\\
\texttt{$\{h_d,h_u,s\}_{_\text{False\_vac.}}$} {\tt (GeV)}  & $-$    & $\{0,~0,661.7\}$ & $\{0,~0,662.3\}$  \\
\texttt{$\{h_d,h_u,s\}_{_\text{True\_vac.}}$} {\tt (GeV)}  & $-$    & $\{9.5,31.5,668.2\}$ & $\{12.8,41.6,669.0\}$ \\\hline
%
\texttt{$T_n$ {\tt (GeV)} / (Nucleation) Transition type}  &  $-$  & 96.2 / 1st-order & 55.9 / 1st-order  \\
\texttt{$\{h_d,h_u,s\}_{_\text{False\_vac.}}$} {\tt (GeV)}  & $-$     & $\{0,~0,~0\}$ & $\{0,~0,~0\}$ \\
\texttt{$\{h_d,h_u,s\}_{_\text{True\_vac.}}$} {\tt (GeV)} & $-$ & $\{67.0,197.8,774.8\}$ & $\{68.1, 199.2, 759.2\}$ \\
%
\texttt{$\gamma_{_{\rm{EW}}} = \Delta_{SU(2)}/T_n$}  & $-$   & $2.2$ &  $3.8$ \\ \hline
\texttt{CheckMATE} result  &  Excluded    & Excluded &  Excluded \\
$r$-value              &  1.12      & 1.01 & 2.13 \\
Analysis ID         &  CMS$\_$SUS$\_$16$\_$039~ & CMS$\_$SUS$\_$16$\_$039~  & CMS$\_$SUS$\_$16$\_$039~ \\
Signal region ID &  SR$\_$A30  & SR$\_$A30 &  SR$\_$G05\\[0.05cm]\hline
\end{tabular}
}}
\caption{Benchmark scenarios allowed by all relevant theoretical and experimental constraints except for those from the LHC searches for the electroweakinos.}
\label{tab:BPs_set1}
\end{center}
\end{table}
%
%

In table~\ref{tab:BPs_set1}, we present a few benchmark points that exhibit SFOEWPT but are ruled out by the electroweakino searches at the LHC. 
A dedicated \checkmate~analysis excludes BP-D1 (with $r$=$1.12$) and  BP-D2 (with $r$=$1.01$) via a CMS 
analysis~\cite{CMS:2017moi} of 35.6~fb$^{-1}$ data in the $3\ell +\slashed{E}_T$ final state 
where an opposite-sign, same-flavor lepton ($e$ or $\mu$)-pair originates from the decay of an on-shell $Z$-boson coming from the decay of a heavier higgsino-like neutralino. 
Latest LHC analyses for the same final state which exploit 139~\fbinv~of data would anyway exclude more convincingly (i.e., with a larger $r$-value) such scenarios with  $\mueff$ as small as 275 GeV.
In the final states with more than 3 leptons signal region of the same CMS analysis~\cite{CMS:2017moi} excludes BP-D3 rather emphatically ($r=2.13$).
Note that BP-D2 and BP-D3 benchmark scenarios are cosmologically good points, however BP-D1 is not.  In BP-D1, the system does not nucleate successfully to the true minimum phase and hence it would remain trapped at the metastable false minimum ($\{0,0,0\}$).

%
\begin{table}[tp]
\renewcommand{\arraystretch}{1.13}
\begin{center}
{\tiny\fontsize{6.9}{6.4}\selectfont{
\begin{tabular}{|c|@{\hspace{0.06cm}}c@{\hspace{0.06cm}}|@{\hspace{0.06cm}}c@{\hspace{0.06cm}}|@{\hspace{0.06cm}}c@{\hspace{0.06cm}}|c|@{\hspace{0.06cm}}c@{\hspace{0.06cm}}|}
\hline
\Tstrut
Input/Observables & {BP-A1} & {BP-A2} & {BP-A3} & {BP-A4}  \\
\hline
\Tstrut
$\lambda, \, \kappa, \, \tan\beta$ &  0.609, \, 0.326, \, 1.98  &  0.609, 0.326, \, 1.98 & 0.633, \, 0.216, \, 1.79 & 0.523, \, 0.041, \, 3.65\\
$\alambda, \akappa, \mueff, M_1$~(GeV)  &  477.0, 38.7, 421.8, 480.1 &  477.0, 37.8, 421.8, -365.1 & -558.7, 46.3, -398.7, 286.3 & -1253.9, 138.1, -334.5, -143.8 \\
$m_{\widetilde{Q}_3}, \, m_{\widetilde{U}_3}, \, A_{t}$~(GeV) &  4262.7, 3450.4, -639.2 &  4262.7, 3450.4, -639.2  & 3950.3, 3544.4, 1372 & 2292, 3435.8, 3862.4\\
\hline
\Tstrut
$m_{\chi_{1,2,3,4}^0, \chi_1^\pm}$~(GeV)  & 396, -446, 477, 510, 432 & -361, 415, -448, 493, 432 & 285, -290, -422, -427, -412 & -61, -139, -359, 360, -345 \\
$m_{h_1}, \, m_{h_2}, \, m_{a_1}, \, m_{H^{\pm}}$~(GeV) & 122.6, 449.2, 75.01 816.5 & 122.7, 449.0, 79.0, 821.4 & 126.9, 288.5, 84.8, 800.9 & 74.0, 124.7, 121.0, 1293.3 \\
\hline
\Tstrut
$\Omega h^2$  &  $3.78 \times 10^{-4}$  &  0.107  & 0.119 & $1.96 \times 10^{-3}$  \\[0.10cm]
$\sigma^{\rm SI}_{\chi^0_1-p(n)}\times \xi$~(cm$^2$)  &    $1.2(1.3)\times 10^{-46}$&  $7.2(7.6)\times 10^{-48}$  & $1.2(1.2)\times 10^{-46}$ & $4.1(4.3)\times 10^{-47}$   \\[0.30cm]
$\sigma^{\rm SD}_{\chi^0_1-p(n)}\times \xi$~(cm$^2$)  &    $4.6(4.5)\times 10^{-44}$&  $9.4(7.3)\times 10^{-42}$  & $3.5(2.8)\times 10^{-42}$  & $1.1(0.8)\times 10^{-41}$   \\[0.15cm]
\hline
\texttt{First $T_n$ {\tt(GeV)}}  &  946.6 / 1st-order    & 945.6 / 1st-order & 644.3 / 1st-order & 116.9 / 1st-order\\
\texttt{$\{h_d,h_u,s\}_{_\text{False\_vac.}}$} {\tt (GeV)} & $\{0,~0,~0\}$  &  $\{0,~0,~0\}$ &  $\{0,~0,~0\}$ &  $\{0,~0,~0\}$ \\
\texttt{$\{h_d,h_u,s\}_{_\text{True\_vac.}}$} {\tt (GeV)}  & $\{0,~0,64.9 \}$   & $$\{0,~0, 66.2 \}$ $ & $\{0,0,-104.8\}$ & $\{30.3,113.8,-877.4\}$ \\
%
\hline
\texttt{Second $T_n$ {\tt (GeV)}}  & 90.2 / 1st-order  & 86.2 / 1st-order & 94.5 / 1st-order & $-$\\
\texttt{$\{h_d,h_u,s\}_{_\text{False\_vac.}}$} {\tt (GeV)}  & $\{0,0,1000.9\}$    & $\{0,0,1000.8\}$  & $\{0,0,-914.9\}$ & $-$\\
\texttt{$\{h_d,h_u,s\}_{_\text{True\_vac.}}$} {\tt (GeV)}  & $\{44.2,86.9,1000.6\}$    & $\{57.1,112.5,1000.3\}$ & $\{48.5,85.6,-914.8\}$ & $-$\\
\hline
\texttt{$\gamma_{_{\rm{EW}}} = \Delta_{SU(2)}/T_n$}  & 1.08   & 1.46 &  1.04 & 1.01\\\hline
\Tstrut
\texttt{CheckMATE} result  &    Allowed  &  Allowed  &  Allowed  & Allowed  \\
$r$-value               &   0.03 &  0.08     &  0.14  & 0.55  \\
Analysis ID          &   CMS$\_$SUS$\_$16$\_$039 &  CMS$\_$SUS$\_$16$\_$039 &   CMS$\_$SUS$\_$16$\_$039 & CMS$\_$SUS$\_$16$\_$039\\
Signal region ID &    SR$\_$A01 &  SR$\_$A08  & SR$\_$A28  & SR$\_$A31\\[0.05cm]\hline
\end{tabular}
}}
\caption{Benchmark scenarios (with successful nucleation) allowed by all relevant theoretical and experimental constraints including the recent ones from the LHC searches for the electroweakinos.}
\label{tab:BPs_set2}
\end{center}
\end{table}
Thus, table~\ref{tab:BPs_set1} indicates how different 
types of spectra for the light higgsino-like electroweakinos (i.e., smaller
$\mueff$), which aid SFOEWPT and are allowed by all relevant constraints including those coming from the DM sector, get excluded by the LHC analyses with only $\sim 36$ \fbinv~of data even when the latter's sensitivities to the targeted final states deteriorate significantly for the kind of spectra in use. 
We choose our benchmark scenarios in such a way that those end up with $r \gtrsim 1$.
Such `r'-value reflects how relatively light electroweakinos are still allowed before they start attracting bounds from the LHC analyses.
Of course, the latest LHC analyses with 139 \fbinv~of data are likely
to increase these mass-bounds ($\mueff$) but have not yet been included in the recast package.


In table~\ref{tab:BPs_set2}, we present a few benchmark scenarios that exhibit an SFOEWPT but now also pass the latest constraints coming from electroweakino searches at the LHC as well.
This pushes $\mueff$ up which impedes
 a successful SFOEWPT from nucleating.
The SFOEWPT now prefers to follow a two-step pattern.
This is common when the trivial and global minima in the field space are widely separated which is the case here given that a larger value for $\mueff$ corresponds to a larger $\vs$ at $T = 0$ for a given value of $\lambda$, which is a characteristic that regulates the field separation at $T_c$. In the dark sector, it is also found that to comply with the latest DMDD constraints, the bino/singlino-like LSP must satisfy various DMDD blind-spot conditions~\cite{Abdallah:2020yag}. 
\begin{figure}[tp]
\begin{center}
\includegraphics[height=4.7cm,width=0.39\linewidth]{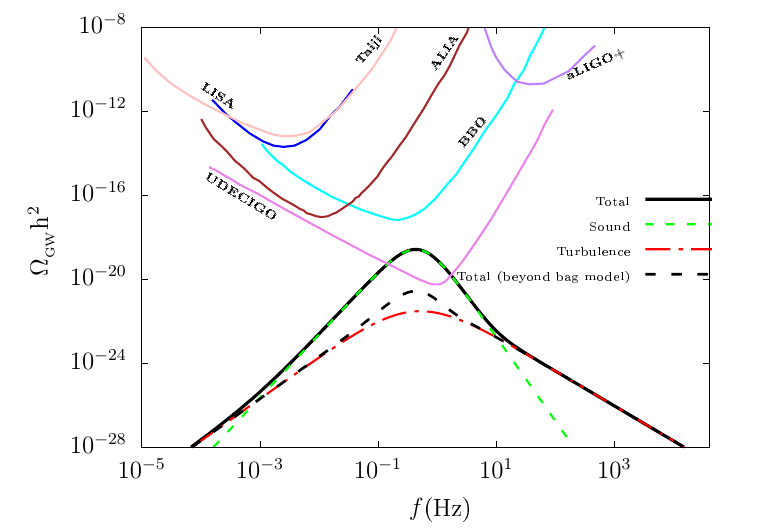}
\hskip 10pt
\includegraphics[height=4.7cm,width=0.39\linewidth]{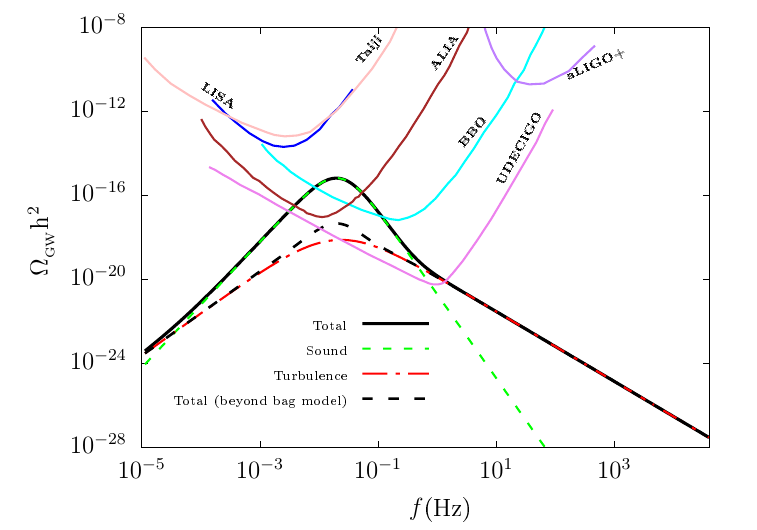}
\hskip 10pt
\includegraphics[height=4.7cm,width=0.39\linewidth]{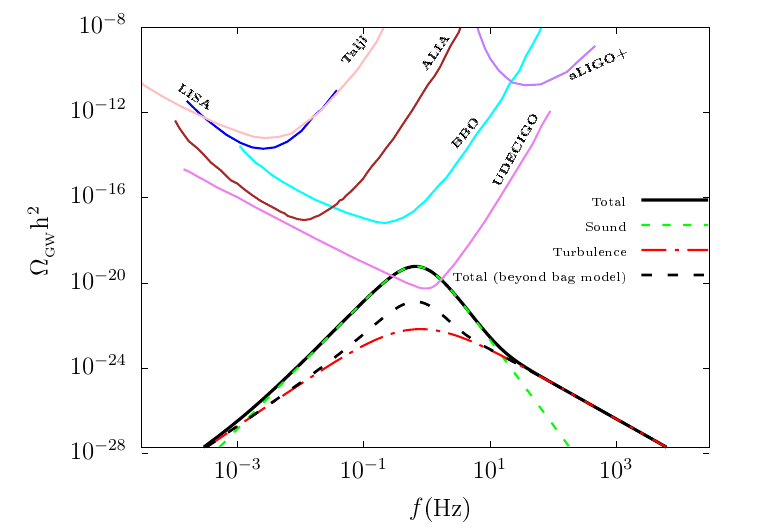}
\hskip 10pt
\includegraphics[height=4.7cm,width=0.39\linewidth]{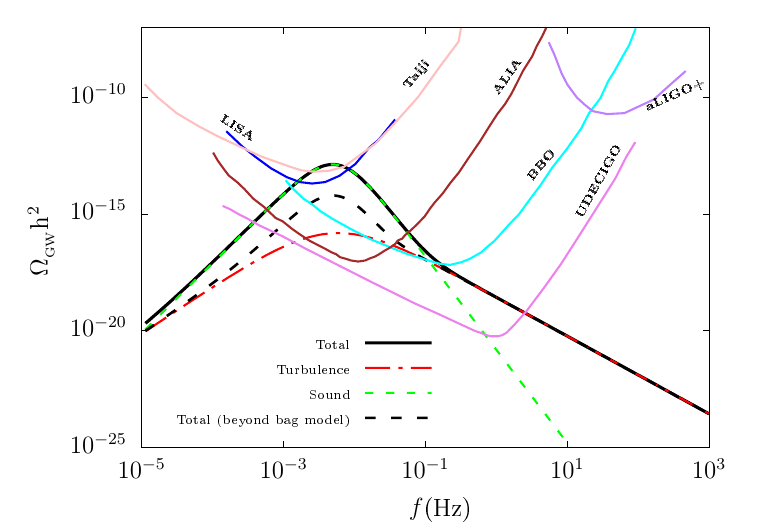}~~
\caption{GW energy density spectrum for the 
benchmark scenarios BP-A1 (top, left), BP-A2 (top, right), BP-A3 (bottom, left) 
and BP-A4 (bottom, right) illustrated against the detection sensitivities 
of some GW experiments like LISA, TianQin, aLigo+, Taiji, ALIA, BBO and
U-DECIGO.}
\label{fig:GW-freq-plot}
\end{center}
\vspace{-0.5cm}
\end{figure}
%

The GW (frequency) spectra for these benchmark points BP-A1 to
BP-A4 are presented in 
figure~\ref{fig:GW-freq-plot}.
 In all these cases, the individual contributions from sound waves  (within the bag model) and turbulence  are shown with broken lines, in green and red colors, respectively. The total GW spectra within (beyond) the bag model, are denoted by solid (broken) black lines. 
The peak of GW spectrum lies within the sensitivity of various future proposed GW detectors.
For all these benchmark scenarios, the quantity signal-to-noise ratio (SNR) that measures the detectability of the GW signal is way below than the detectability criterion of the LISA experiment. 
\vspace{-0.2cm}
\section{Summary and outlook}
\label{sec:summary}
\vspace{-0.2cm}
The work sheds light on what the recent searches of the electroweakinos at the LHC have to say about the viability of  SFOEWPT in the $Z_3$-NMSSM  while complying with all constraints from various pertinent theoretical and experimental sectors, including the DM sector experimental finding. While an SFOEWPT is favored in the range of a few hundreds of a GeV of $\mueff$, the latest constraints are tending to push $\mueff$ steadily above such a range. The general goal of such a study could then be to check if there is a meeting ground somewhere in the middle where both constraints are simultaneously complied with. 
A middle ground can thus only be found if the reported constraints from the LHC could be evaded under circumstances that have not been considered explicitly by the LHC experiments.
The two-step phase transition is a more likely
phenomenon for $\mueff$ on the larger side with the first transition occurs along the singlet field direction followed
by the second one in the SU(2)-field directions.
Stochastic GW signals originating from  FOEWPT in these scenarios are likely to remain too weak to get detected in the LISA experiment.

New studies on the searches of electroweakinos at relatively higher masses and in difficult scenarios like the compressed ones using the data coming from the recently terminated run~2 phase of LHC and from the HL-LHC would continue to shed light on the motivated region of parameter space for EWBG within the \z3nmssm. A further study is required to estimate the BAU in the allowed
scenarios that have been discussed in this work.
Furthermore, there are several scopes to improve the theoretical calculations of bubble wall profile, $CP$-violation, solving transport equations, GW spectrum, etc. 
which are important for the accurate estimation of the
EWBG preferred region of NMSSM parameter space.
A complementarity/interplay search among  GW, DM and LHC physics is likely to be rather intriguing and we reserve it for future works.

\end{document}